\documentclass[sn-mathphys]{sn-jnl}

\jyear{2021}%

\theoremstyle{thmstyleone}%
\theoremstyle{thmstyletwo}%
\theoremstyle{thmstylethree}%

\usepackage{subcaption}
\usepackage{booktabs, multirow}
\usepackage{tabularx}

\usepackage{array}
\usepackage{amssymb}
\usepackage{amsmath}
\usepackage{epsfig}
\usepackage{graphicx}
\usepackage{amsmath}
\usepackage{amssymb}
\usepackage{subcaption}
\usepackage{booktabs}
\usepackage{array}
\usepackage{graphicx}
\usepackage{xcolor}
\usepackage{soul} 
\newcolumntype{P}[1]{>{\centering\arraybackslash}p{#1}}

\newcommand{\ourmethod}[1]{Neuro-DynaStress}

\begin{document}

\title[\ourmethod{}: Predicting \textbf{Dynamic Stress} Distributions]{\ourmethod{}: Predicting \textbf{Dynamic Stress} Distributions in Structural Components}


\author*[1,2]{\fnm{Hamed} \sur{Bolandi}}\email{bolandih@msu.edu}

\author[2]{\fnm{Gautam} \sur{Sreekumar}}\email{sreekum1@msu.edu}

\author[1,2]{\fnm{Xuyang} \sur{Li}}\email{lixuyan1@msu.edu}

\author[1]{\fnm{Nizar} \sur{Lajnef}}\email{lajnefni@msu.edu}

\author[2]{\fnm{Vishnu Naresh} \sur{Boddeti}}\email{vishnu@msu.edu}

\affil*[1]{\orgdiv{Civil and Environmental Engineering}, \orgname{Michigan State University}, \orgaddress{\street{Shaw Lane}, \city{East Lansing}, \postcode{48824}, \state{MI}, \country{USA}}}

\affil[2]{\orgdiv{Computer Science and Engineering}, \orgname{Michigan State University}, \orgaddress{\street{Shaw Lane}, \city{East Lansing}, \postcode{48824}, \state{MI}, \country{USA}}}


\abstract{ Structural components are typically exposed to dynamic loading, such as earthquakes, wind, and explosions. Structural engineers should be able to conduct real-time analysis in the aftermath or during extreme disaster events requiring immediate corrections to avoid fatal failures. As a result, it is crucial to predict dynamic stress distributions during highly disruptive events in real time. Currently available high-fidelity methods, such as Finite Element Models (FEMs), suffer from their inherent high complexity and are computationally prohibitive. Therefore, to reduce computational cost while preserving accuracy, a deep learning model, \ourmethod{}, is proposed to predict the entire sequence of stress distribution based on finite element simulations using a partial differential equation (PDE) solver.  The model was designed and trained to use the geometry, boundary conditions and sequence of loads as input and predict the sequences of  high-resolution stress contours. The proposed framework's performance is compared to finite element simulations using a PDE solver.}


\maketitle

\section{Introduction}\label{sec1}

Numerical analysis methods, such as Finite Element Analysis~(FEA), are typically used to conduct stress analysis of various structures and systems for which it is impractical or hard to determine an analytical solution. Researchers commonly use FEA methods to evaluate the design, safety and maintenance of different structures in various fields, including aerospace, automotive, architecture and civil structural systems. The current workflow for FEA applications includes: (i) modeling the geometry and its components, (ii) specifying material properties, boundary conditions, meshing, and loading, (iii) dynamic analysis, which may be time-consuming based on the complexity of the model.  The time requirement constraint and the complexity of the current FEA workflow make it impractical for real-time or near real-time applications, such as in the aftermath of a disaster or during extreme disruptive events that require immediate corrections to avoid catastrophic failures.\\

\begin{figure}[t]
    \centering
    \includegraphics[width=0.95\textwidth]{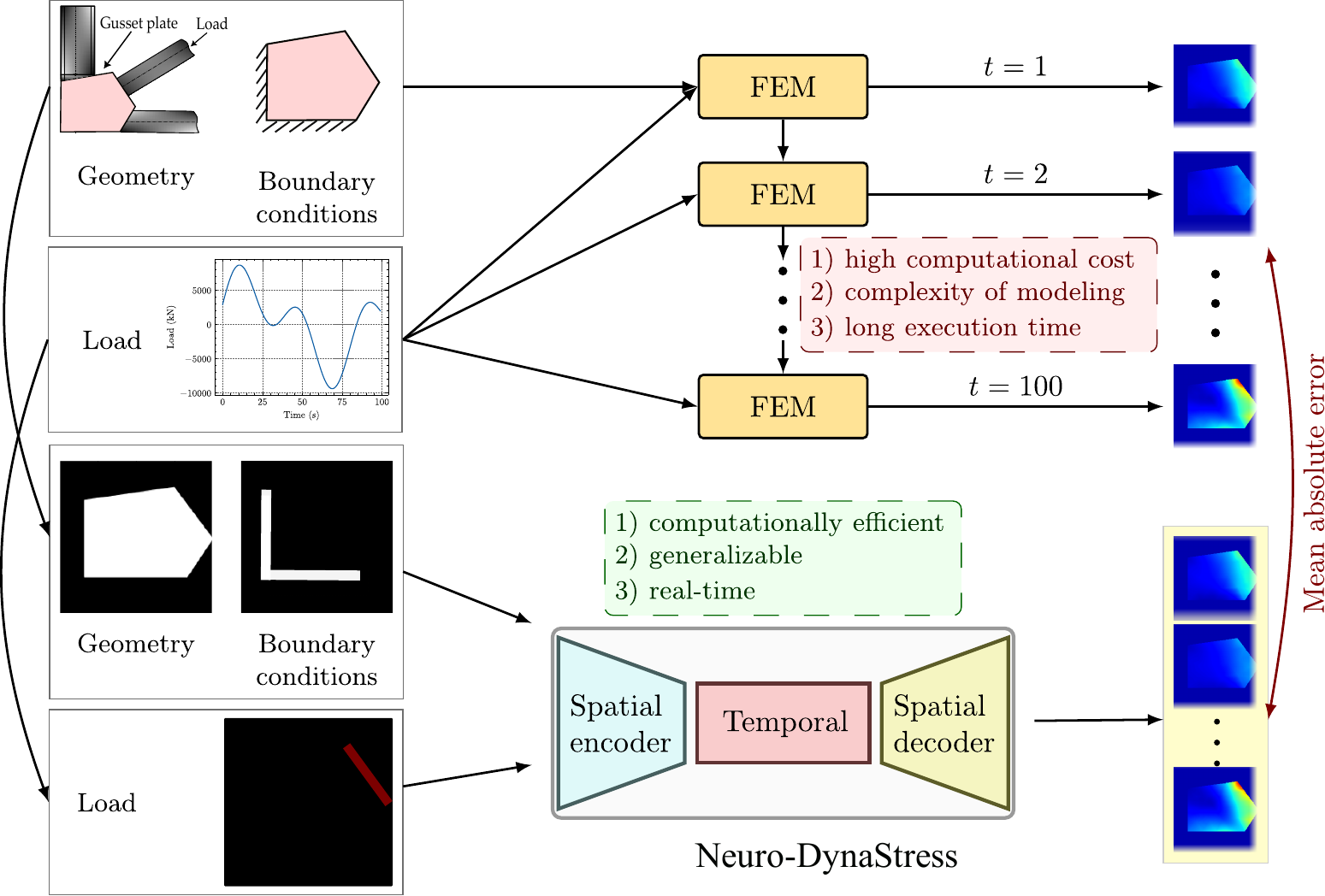}
    \caption{{\textbf{Overview:} Unlike FEM, our proposed \ourmethod{} is computationally efficient and facilitates real-time analysis. The existing workflow for FEM applications includes: (i) modeling the geometry and its components, (ii) specifying material properties, boundary conditions, meshing, and loading, (iii) dynamic analysis, which may be time-consuming based on the complexity of the model. Our \ourmethod{} takes geometry, boundary condition, and load as input and predicts the dynamic stress distribution at all time steps in one shot.} 
    \label{fig:overview}}
\end{figure}

Based on the steps of FEA described above, performing a complete stress analysis with conventional FEA has a high computational cost. In order to overcome this problem, some recent works have proposed deep neural network~(DNN)-based methods to predict stress distributions in both intact and damaged structural components~\cite{bolandi2022a,bolandi2022deep}, bypassing the need for static finite element analysis. But these works are not suitable for dynamic finite element analysis. We propose an architecture that can act as a surrogate for FEA solvers for dynamic FEA while avoiding the computational bottlenecks involved. To demonstrate its utility, we model the stress distribution in gusset plates under dynamic loading. Bridges and buildings rely heavily on gusset plates as one of their most critical components. Gusset plates are designed to withstand lateral loads such as earthquakes and winds, which makes fast dynamic models valuable in avoiding catastrophic failures.\\

The main idea here is to train a model that can later be used when real-time estimations are needed, such as in the aftermath of extreme disruptive events. For example, focusing on critical structural components, there is a need for immediate assessment following a disaster or during extremely disruptive events to guide corrective actions. Engineers could rely on the proposed computationally efficient algorithms to determine stress distributions over damaged gusset plates and apply the proper rehabilitation actions. They need to be able to analyze gusset plates quickly and accurately, which is what our model can provide. To our knowledge, this work is the first to predict dynamic stress distribution in the specific domain of steel plates.

\section{Related Work}\label{sec2}
The most recent works in data-driven applications of scientific machine learning have included design and topology optimization~\cite{umetani2017exploring,yu2019deep}, data-driven approaches in fluid dynamics~\cite{farimani2017deep,kim2019deep}, molecular dynamics simulation~\cite{goh2017deep,mardt2018vampnets}, and material properties prediction~\cite{mohammadi2014multigene,sarveghadi2019development,mousavi2012new,bolandi2019intelligent}. Atalla et al.~\cite{atalla1998model} and Levin et al.~\cite{levin1998dynamic} have used neural regression for FEA model updating. More recently, DL has shown promise in solving traditional mechanics problems. Some researchers used DL for structural damage detection, a promising alternative to conventional structural health monitoring methods~\cite{fan2018automatic,dung2019autonomous}.

Javadi et al.~\cite{javadi2003neural} used a typical neural network in FEA as a surrogate for the traditional constitutive material model. They simplified the geometry into a feature vector which approaches hard to generalize complicated cases. The numerical quadrature of the element stiffness matrix in the FEA on a per-element basis was optimized by Oishi et al.~\cite{oishi2017computational} using deep learning. Their approach helps to accelerate the calculation of the element stiffness matrix. Convolutional Neural Networks (CNN) are commonly used in tasks involving 2D information due to the design of their architecture. Recently, Madani et al.~\cite{madani2019bridging} developed a CNN architecture for stress prediction of arterial walls in atherosclerosis. Also, Liang et al.~\cite{liang2018deep} proposed a CNN model for aortic wall stress prediction. Their method is expected to allow real-time stress analysis of human organs for a wide range of clinical applications.

Gulgec et al.~\cite{gulgec2019convolutional} proposed a CNN architecture to classify simulated damaged and intact samples and localize the damage in steel gusset plates. Modares et al.~\cite{modarres2018convolutional} conducted a study on composite materials to identify the presence and type of structural damage using CNNs. Also, in order to detect concrete cracks without calculating the defect features, Cha et al.~\cite{cha2017deep} proposed a vision-based method based on convolutional neural networks (CNNs). Do et al.~\cite{do2019fast} proposed a method for forecasting the crack propagation in risk assessment of engineering structures based on “long short-term memory” and “multi-layer neural network”. An approach for predicting stress distribution on all layers of non-uniform 3D parts was presented by Khadilkar et al.~\cite{khadilkar2019deep}. More recently, Nie et al.~\cite{nie2020stress} developed a CNN-based method to predict the low-resolution stress field in a 2D linear cantilever beam. Jiang et al.~\cite{jiang2021stressgan} developed a conditional generative adversarial network for low-resolution von Mises stress distribution prediction in solid structures.

Some studies have been conducted to develop methods of predicting structural response using ML models. Dong et al.~\cite{yinfeng2008nonlinear} proposed a support vector machine approach to predict nonlinear structural responses. Wu et al.~\cite{wu2019deep} Utilized deep convolutional neural networks to estimate the structural dynamic responses. Long short-term memory (LSTM)~\cite{hochreiter1997long} was used by Zhang et al.~\cite{zhang2019deep} to predict nonlinear structural response under earthquake loading. Fang et al.~\cite{fang2022combined} proposed a deep-learning-based structural health monitoring (SHM) framework capable of predicting a dam's structural dynamic responses once explosions are experienced using LSTM. Kohar et al.~\cite{kohar2021machine} used 3D-CNN-autoencoder and LSTM to predict the force-displacement response and deformation of the mesh in vehicle crash-worthiness. Schwarzer et al.~\cite{schwarzer2019learning} construct a neural network architecture that combines a graph convolutional neural network (GCN) with a recurrent neural network (RNN) to predict fracture propagation in brittle materials. Lazzara et al.~\cite{lazzara2022surrogate} proposed a dual-phase  LSTM Auto-encoder-based surrogate model to predict aircraft dynamic landing response over time. Jahanbakht et al.~\cite{jahanbakht2022sediment} presented an FEA-inspired DNN using an attention transformer to predict the sediment distribution in the wide coral reef.

The few models that studied stress predictions suffer from the problem of low-resolution predictions, making them unsuitable for decision-making after a catastrophic failure. To the best of our knowledge, this is the first work to predict dynamic stress distribution in the specific domain of steel plates with high accuracy and low latency. The algorithm takes the geometry, boundary conditions, and time histories as input and renders the dynamic von Mises stress distribution as an output. We modeled the steel plates as gusset plates with dynamic loading applied at different edges, different boundary conditions, and varying complex geometries.

\section{Methods}\label{sec3}
\subsection{Data Generation}\label{subsec2}

\begin{figure}[!h]
    \centering
    \includegraphics[width=0.3\textwidth]{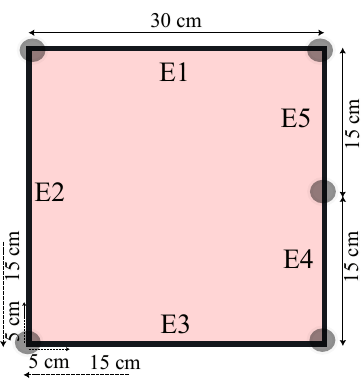}
    \caption{Basic schematic topology for initializing the steel plate geometries.\label{fig: schematic topology}}
\end{figure}

\begin{figure}[!h]
    \centering
    \includegraphics[width=0.95\textwidth]{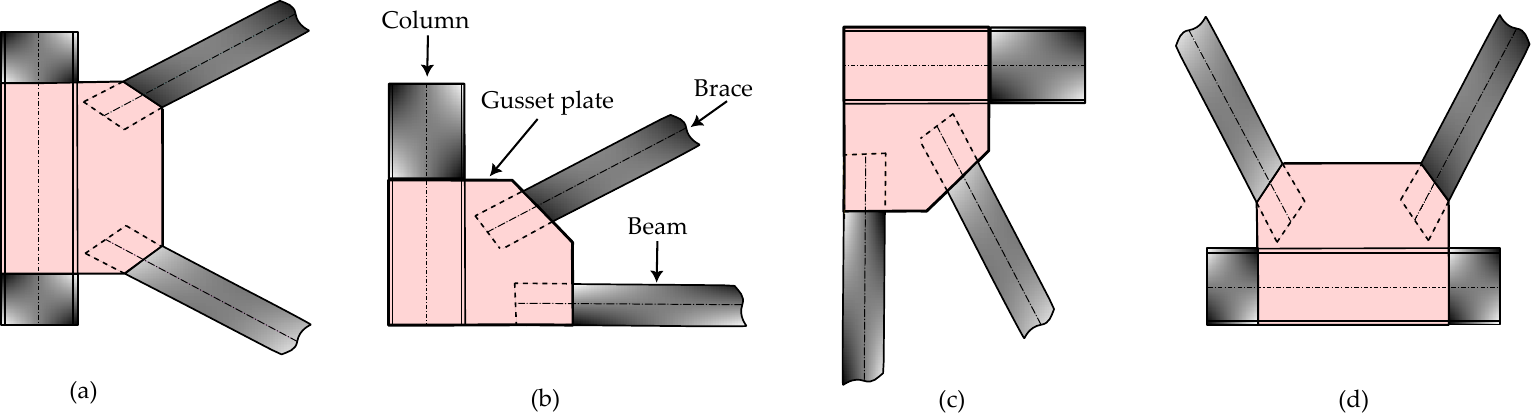}
    \caption{Some of the most common gusset plates in practice.\label{fig:common gussets}}    
\end{figure}

Two-dimensional steel plate structures with five edges, E1 to E5 denoting edges 1 to 5, as shown in Fig.~\ref{fig: schematic topology}, are considered homogeneous and isotropic linear elastic materials. Various geometries are generated by changing the position of each node in horizontal and vertical directions, as shown in Fig.~\ref{fig: schematic topology}, which led to 1024 unique pentagons. The material properties remain unchanged, isotropic for all samples. The 2D steel plates approach the geometry of gusset plates. Gusset plates connect beams and columns to braces in steel structures. The behavior and analysis of these components are critical since various reports have observed failures of gusset plates subject to lateral loads~\cite{zahraei2007destructive,zahrai2014towards,zahrai2019numerical,bolandi2013influence}. The boundary conditions and time-history load cases are considered to simulate similar conditions in common gusset plate structures under external loading. Some of the most common gusset plates configurations in practice are shown in Fig.~\ref{fig:common gussets}.

\begin{figure}[!h]
    \centering
    \includegraphics[width=0.95\textwidth]{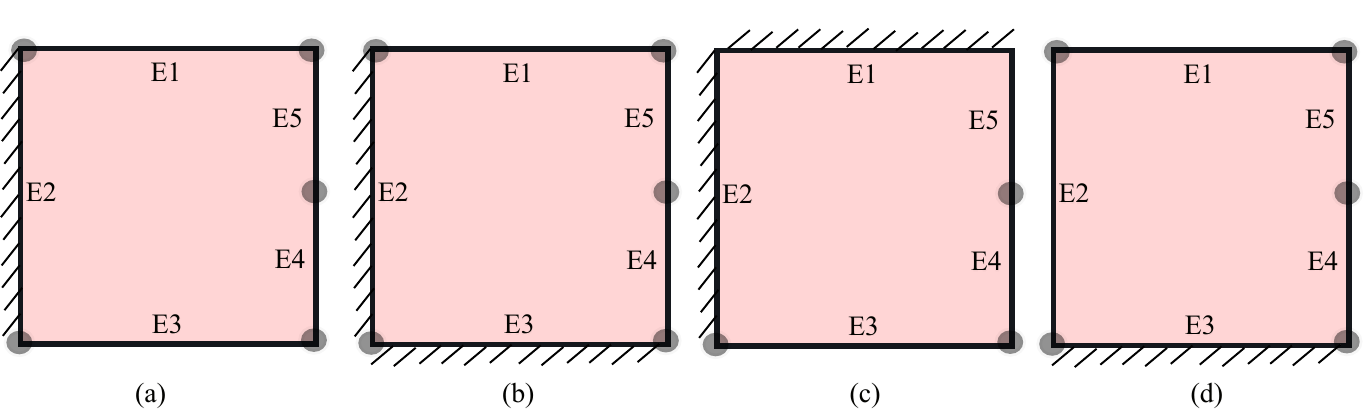}
    \caption{Different types of boundary conditions for initializing population.\label{fig:boundary conditions}}    
\end{figure}

\begin{figure}[!h]
\centering
\subcaptionbox{\label{fig:hist1}}{\includegraphics[width=\textwidth]{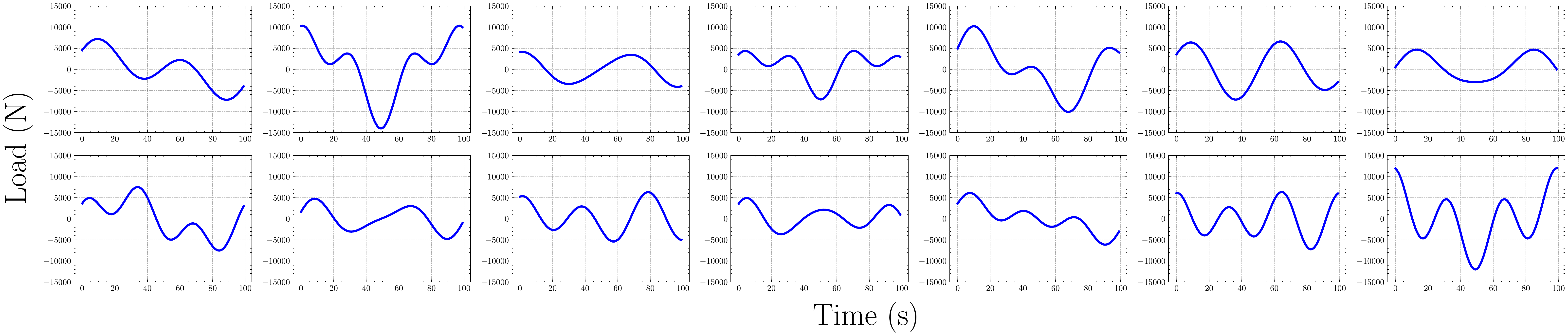}}
\subcaptionbox{\label{fig:hist2}}{\includegraphics[width=\textwidth]{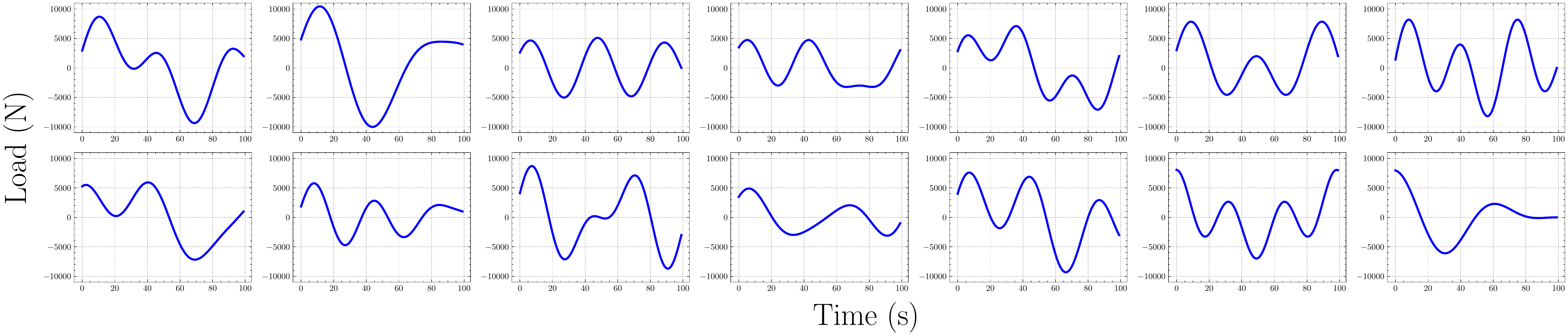}}
\caption{Time histories (a) Horizontal direction (b) Vertical direction\label{fig:time histories}}
\end{figure}

A total of 57,344 unique samples were created by combining 14 random time-history load cases and four most common boundary conditions in gusset plates. Boundary conditions are shown in Fig.~\ref{fig:boundary conditions}, mimicking the real gusset plates’ boundary conditions. All the translation and rotational displacements were fixed at the boundary conditions. The range for width and height of the plates is from 30 cm to 60 cm. Each time history consists of 100 time steps generated with random sine and cosine frequencies. The frequencies range between 1 and 3 HZ, with amplitudes ranging from 2 to 10 kN at intervals of 2 kN. All time histories in horizontal and vertical directions are shown in Fig.~\ref{fig:time histories}. Considering 100 time steps, each interval is 0.01 seconds, making the total time equal to 1 second. All the details for the input variables used to initialize the population are shown in Table~\ref{table:input-variables}.

\setlength{\tabcolsep}{4pt}
\begin{table}[!h]
\begin{center}
\caption{Input variable}
\label{table:input-variables}
\begin{tabular}{m{1.5cm} >{\centering}m{1.5cm} >{\centering}m{1.3cm} >{\centering}m{2cm} >{\centering}m{1.cm} >{\centering}m{1cm} >{\centering\arraybackslash}m{1.3cm}}
\toprule
 Geometry & Boundary conditions & Load position & Frequencies (HZ) & Load (kN) & Time steps & Total time (s)\\
\noalign{\smallskip}
\midrule
\noalign{\smallskip}
 pentagon & E2 & E4E5 & 1,1.5,2,2.5,3 & 2,4,6,8,10 & 100 & 1 \\
 pentagon & E2E3 & E5 & 1,1.5,2,2.5,3 & 2,4,6,8,10 & 100 & 1  \\
 pentagon & E1E2 & E4 & 1,1.5,2,2.5,3 & 2,4,6,8,10 & 100 & 1  \\
 pentagon & E3 & E2E5 & 1,1.5,2,2.5,3 & 2,4,6,8,10 & 100 & 1 \\
\bottomrule
\end{tabular}
\end{center}
\end{table}
\setlength{\tabcolsep}{1.4pt}

\subsection{Input Data}
\begin{figure}[h]
    \centering
    \includegraphics[width=0.95\textwidth]{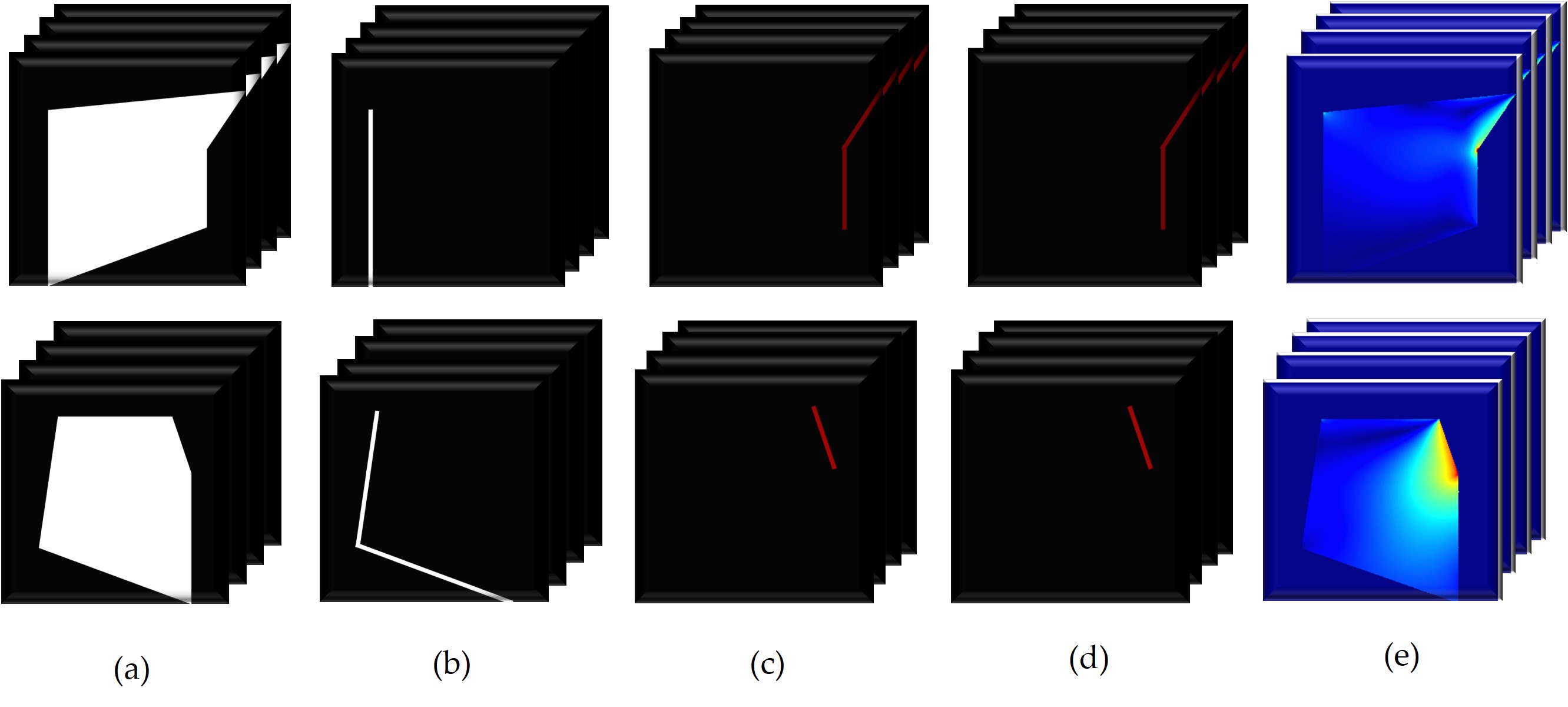}
    \caption{Input and output representation for stress distribution prediction: (a) geometry, (b) boundary condition, (c) horizontal load, (d) vertical load, (e) output\label{fig:Input and output}}    
\end{figure}

The geometry is encoded as a $200\times200$ matrix and, incidentally, a binary image. 0 (black) and 1 (white) denote outside and inside of the geometry, as shown in Fig.~\ref{fig:Input and output}(a). The boundary condition is also represented by another $200\times 200$ pixel binary image, where the constrained edges are defined by 1 (white)  as shown in Fig.~\ref{fig:Input and output}(b). Moreover, each time step of time histories for horizontal and vertical components is encoded in the load position of the corresponding frame. Load positions in each time step have values between 0 and 1, corresponding to each time step of time histories, and all remaining elements are zero. All the load frames of each sample in horizontal and vertical directions are saved as tensors of dimension $100\times 200\times 200$. Figs.~\ref{fig:Input and output}(c) and \ref{fig:Input and output}(d) show loads in the horizontal and vertical directions. The colored load positions in Figs.~\ref{fig:Input and output}(c) and \ref{fig:Input and output}(d) are used only for visualization. Each row of Fig.~\ref{fig:Input and output} represents one of the simulated samples. Details of boundary conditions and their load positions are described in Table~\ref{table:input-variables}.

\subsection{Output Data}
\begin{figure}[h]
    \centering
    \includegraphics[width=0.95\textwidth]{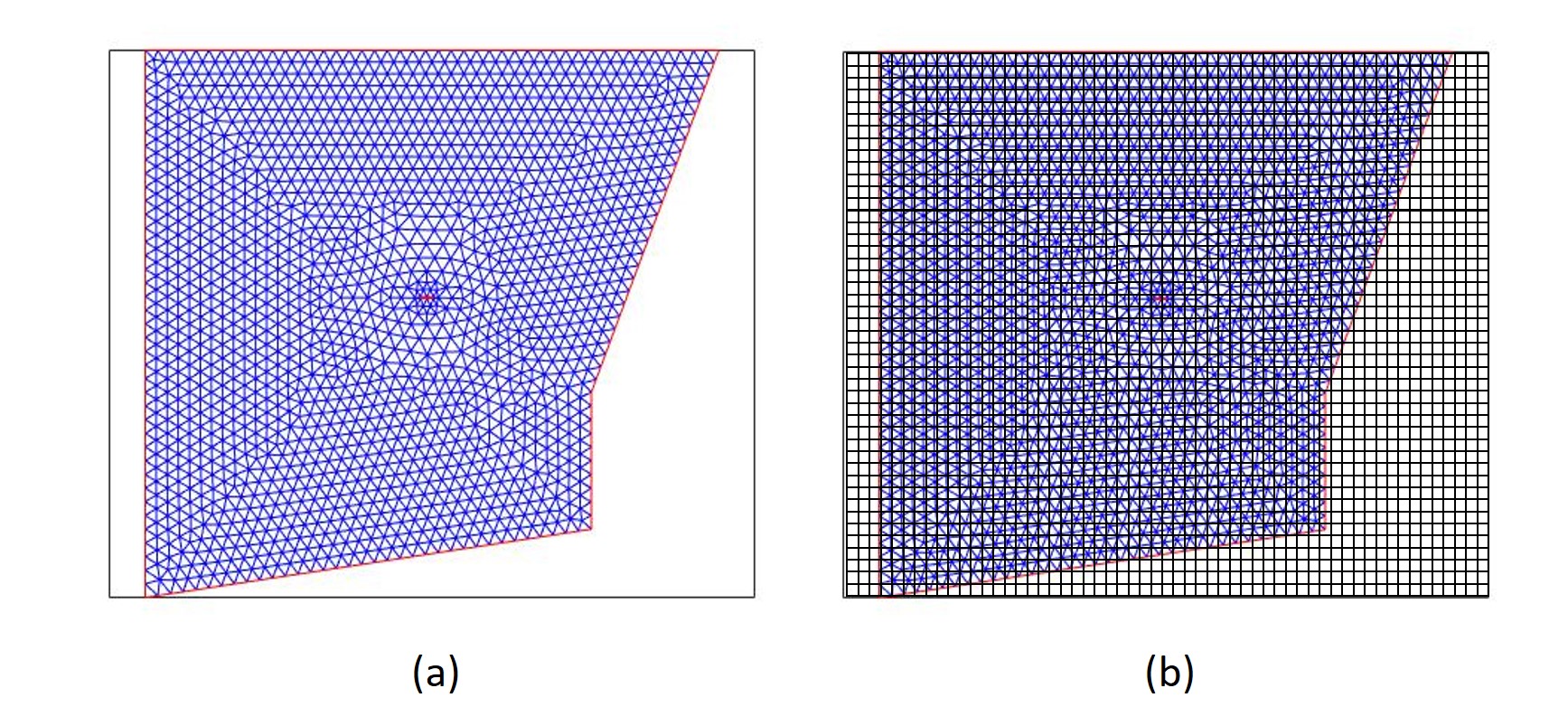}
    \caption{A sample of mesh generation: (a) unstructured triangular mesh, (b) structured gird surface}
    \label{fig:mesh generation}
\end{figure}

FEA was performed using the Partial Differential Equation (PDE) solver in the MATLAB toolbox to obtain the stress distributions of each sample. We used transient-planestress function of MATLAB PDE solver to generate dynamic stress contours as the ground truth of our model. We defined the geometry, boundary condition, material properties and time histories as input and PDE solver returns the sequence of stress distributions corresponding to the inputs. The MATLAB PDE toolbox mesh generator only generates unstructured triangulated meshes incompatible with CNN. The minimum and maximum triangulated mesh sizes are 5 and 10mm, respectively. Since each element should be represented by one pixel in an image, we develop a $200\times200$ grid surface equal to the dimensions of the largest possible geometry. Figs.~\ref{fig:mesh generation}(a) and \ref{fig:mesh generation}(b) show the unstructured mesh and the $200\times200$ grid surface on top of a random sample. The stress values are then interpolated between the triangular elements and grids to determine a stress distribution compatible with our CNN network. The stress values of all the elements outside the material geometry are assigned to zero, as shown in Fig.~\ref{fig:Input and output}(e).

The dimension of the largest sample is $600\times600$ mm, and the smallest is $300\times300$ mm. Using a mesh grid of $200\times200$ on top of samples made each element $3\times3$ mm, which means that each frame of output has 40000 pixels. This high-resolution dataset led to achieving significant accuracy. The maximum and minimum von Mises stress values for elements among the entire dataset are 279,370 and -980 MPa, respectively. We normalized all the output data between 0 and 1 to ensure faster convergence and encoded it to $200\times200$ for each frame.

\subsection{Stress Calculation}
The steps for linear finite element analysis' stress calculation, which is part of phase (iii) of FEA's workflow elaborated in the introduction section, are as follows:

\begin{equation}\label{first_eq}
KQ = F
\end{equation}

where $K$ denotes a global stiffness matrix, $F$ is the load vector applied at each node, and $Q$ denotes the displacement.  A stiffness matrix $K$ consists of elemental stiffness matrices $K_e$:

\begin{equation}\label{second_eq}
K_e = A_e B^{T} D B
\end{equation}

where $B$ represents strain-displacement matrix; $D$ represents stress-strain matrix; and $A_e$ represents area of element. Mesh geometry and material properties determine $B$ and $D$. This will be followed by adding the local stiffness matrix $k_e$ to the global stiffness matrix. The displacement boundary conditions are encoded using the corresponding rows and columns in the global stiffness matrix $K$. Solving $Q$ can be achieved using direct factorization or iterative methods.

As a result of calculating the global displacement using equation ~\ref{first_eq}, we can calculate the nodal displacements $q$ then we can calculate the stress tensors of each element as follows:

\begin{equation}\label{third_eq}
{\sigma} = D B q
\end{equation}

where $\sigma$ specifies the tensor of an element. The 2-D von Mises Stress criterion is then used to calculate each element's von Mises Stress:

\begin{equation}\label{fourth_eq}
{\sigma_{v_m}} = \sqrt{\sigma_x^{2} + \sigma_y^{2} - \sigma_x \sigma_y + 3\tau_{x_y}^{2}}
\end{equation}

where $\sigma_{v_m}$ denotes von Mises Stress, $\sigma_x$, $\sigma_y$ are the normal stress components and $\tau_{x_y}$ is the shear stress component.\\

\section{Proposed Methodology}

\begin{figure}[h]
    \centering
    \includegraphics[width=\textwidth]{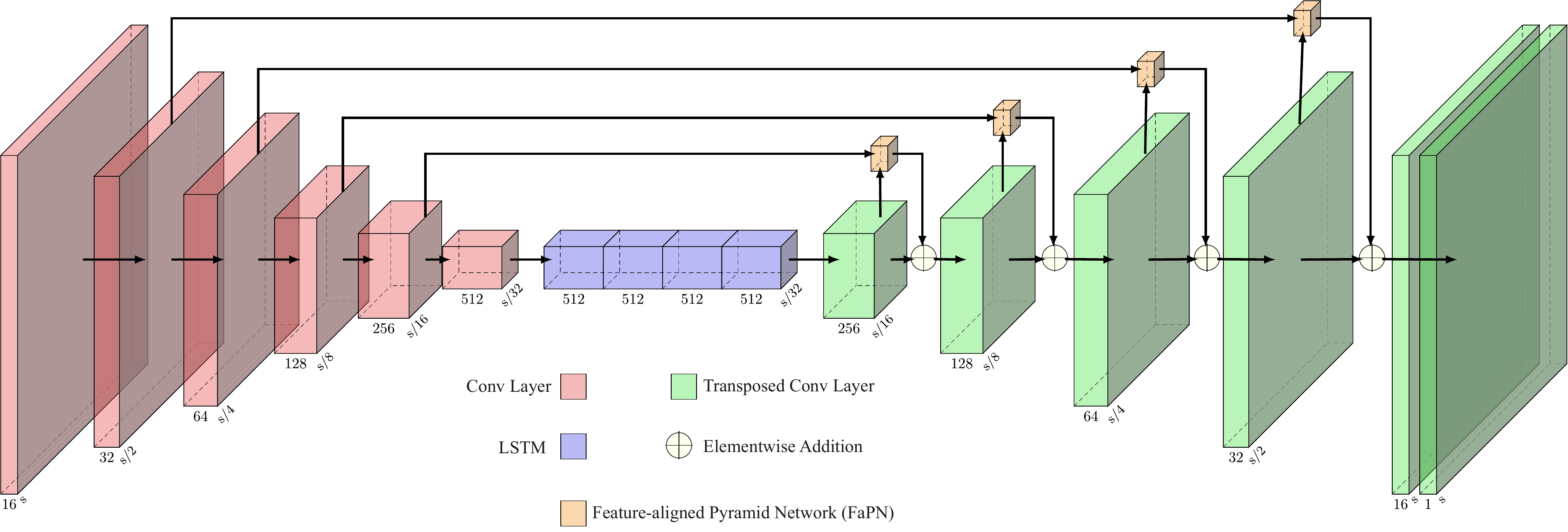}
    \caption{Architecture for the proposed \ourmethod{}. The convolutional encoder maps the raw input data to a latent space. LSTM layers processes the information across different time frames. The final output is obtained from the resulting latent representation using transposed convolutional layers.}
    \label{fig:CNN architecture}
\end{figure}

We use convolutional layers to encode the spatial information from the input. Our hypothesis is that these layers will combine the information in geometry, boundary conditions, and load. A key characteristic of dynamic structural systems is the temporal dependence of their states. LSTM is a suitable architecture for modeling temporal information in sequence and hence is a good choice to model structural dynamic systems in our experiments. For high-quality 2D reconstructions, we use transposed convolutional layers in our decoder. For further improving training and performance, we use modules from the recently proposed feature-aligned pyramid networks~(FaPN)~\cite{huang2021fapn}. FaPN allows the decoder to access information from the encoder directly. Overall, our network architecture consists of four modules: encoder consisting of convolutional layers, temporal module made using LSTM modules, decoder consisting of transposed convolutional layers, and alignment modules acting as connections between encoder and decoder. The number of layers in each module and the number of layers in LSTM modules were chosen based on their performance. The architecture is illustrated schematically in Fig.~\ref{fig:CNN architecture}. The size of layers and hyper-parameters used in the network are summarized in Table~\ref{table:hyperparameters}.

\setlength{\tabcolsep}{4pt}
\begin{table}[!h]
\begin{center}
\caption{Network layers and hyper-parameters}
\label{table:hyperparameters}
 \begin{tabular}{wl{2.5cm} >{\centering}m{2.5cm} >{\centering}m{3cm} >{\centering\arraybackslash}m{3cm} }
\toprule
Type of layers & Number of layers & First layer (H$\times$W$\times$C) &  Last layer (H$\times$W$\times$C)\\
\noalign{\smallskip}
\midrule
\noalign{\smallskip}
Conv &  6 & 200$\times$200$\times$16 &  7$\times$7$\times$512 \\
LSTM &  4 &  1$\times$1$\times$512 &  1$\times$1$\times$512 \\
ConvT &  5 & 13$\times$13$\times$256 &  200$\times$200$\times$16\\
FaPN &  4 & 13$\times$13$\times$256 &  100$\times$100$\times$32 \\
\midrule
\noalign{\smallskip}
 Batch size & Learning rate & Weight decay &  Loss function\\
\midrule
8 & $10^{-4}$ & $10^{-5}$  & MAE\\
\bottomrule
\end{tabular}
\end{center}
\end{table}
\setlength{\tabcolsep}{1.4pt}

\section{Loss Function and Performance Metrics}

We use Mean Absolute Error~(MAE), defined in Eq.~\ref{eq_6} as the primary training loss and metric. To ensure that we do not overfit to a single metric, we also use Mean Relative Percentage Error~(MRPE) to evaluate the overall quality of predicted stress distribution.

\begin{equation}\label{eq_6}
\text{MAE} = \frac{1}{NT} \sum_{N,T}^{n,t} \lvert S(n,t)-\hat{S}(n,t)\rvert\
\end{equation}

\begin{equation}\label{eq_7}
\text{MRPE} = \frac{\text{MAE}}{ \max \(\lvert S(n,t),\hat{S}(n,t)\rvert\)} \times 100
\end{equation}

\noindent where $S(n,t)$ is the true stress value at a node $n$ at time step $t$, as computed by FEA, and $\hat{S}(n,t)$ is the corresponding stress value predicted by our model, $N$ is the total number of mesh nodes in each frame of a sample, and $T$ is a total number of time steps in each sample. As mentioned earlier, we set $T=100$ in our experiments.


\section{Implementation and Computational Performance}
We implemented our model using PyTorch~\cite{paszke2019pytorch} and PyTorch Lightning. AdamW optimizer~\cite{loshchilov2017decoupled} was used as the optimizer with a learning rate of $10^{-4}$. We found that a batch size of 8 gave the best results. The computational performance of the model was evaluated on an AMD EPYC 7313 16-core processor and two NVIDIA A6000 48G GPUs. The time required during the training phase for a single sample with 100 frames and a batch size of 8 was 10 seconds. In the training phase, one forward and backward pass was considered. The inference time for one sample was less than 5 ms which can be considered a real-time requirement. The most powerful FE solvers take between 10 minutes to an hour to solve the same. Therefore, \ourmethod{} is about $72\times10^{4}$ times faster than conventional FE solvers. We consider the minimum time for all processes of modeling geometry, meshing, and analysis of one sample in FE solver to be about 10 minutes. MATLAB PDE solver does not use GPU acceleration. This demonstrates that our proposed approach can achieve the real-time requirement during the  validating phase.

\section{Results and Discussions}
\subsection{Quantitative Evaluation}
Our model is trained on the training dataset for 45 epochs and evaluated on the validation dataset using separate metrics. The training dataset consisted of 48,755, while the validating dataset contained 8,589 samples, together forming the 80\%-20\% split of the whole dataset. The model predicts five frames of output from a sequence of five previous inputs until all 100 frames are predicted. The best validation performance was obtained when we sequenced five frames during validation. The best checkpoint during validation, at epoch 40, is the basis for all error metrics. MRPE for the validating dataset is just 2.3\%.

\subsection{Qualitative Evaluation}
\begin{figure}[!h]
    \centering
    \subcaptionbox{\label{fig:successful predtions1}}{\includegraphics[width=0.95\textwidth]{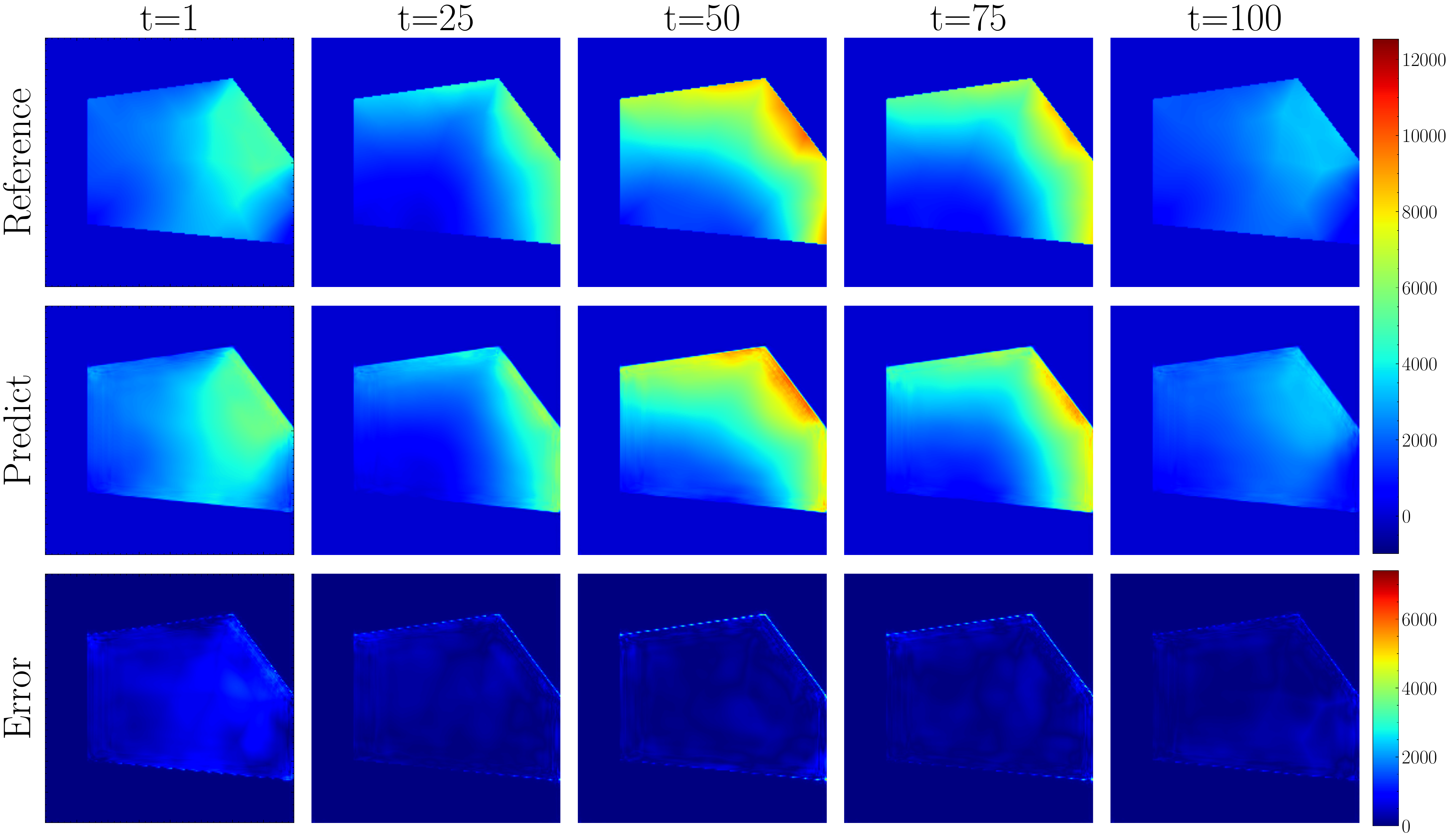}}
    \subcaptionbox{\label{fig:successful predtions2}}{\includegraphics[width=0.95\textwidth]{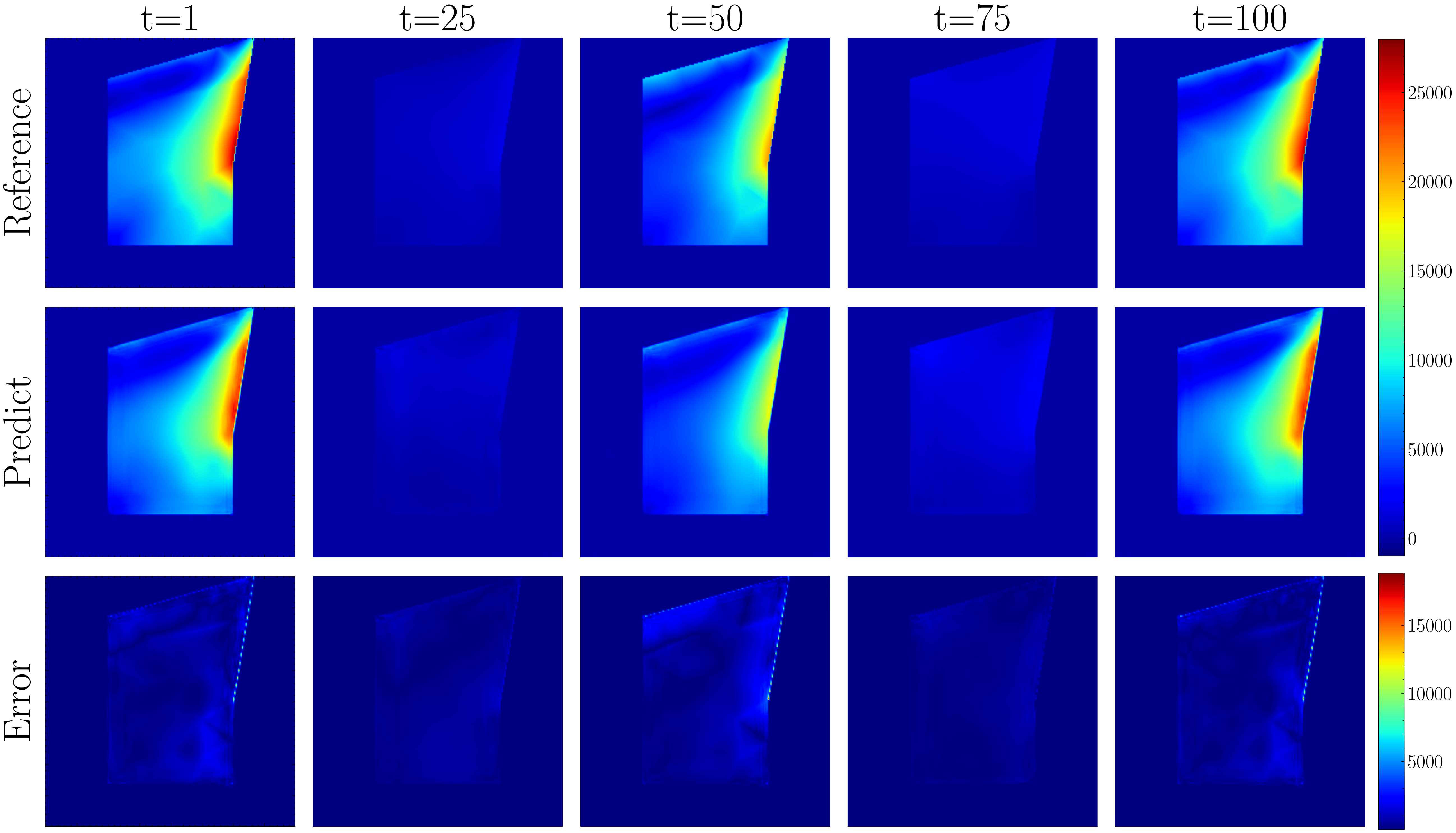}}
    \caption{Successful predicted dynamic stress distribution and their corresponding errors in different time sequences for two samples. The top row corresponds to reference frames and the middle row shows the predictions. The bottom row shows the absolute error between corresponding frames (Unit = MPa)}
\end{figure}

The prediction results for a few randomly selected samples from the validation dataset are visualized in Figs.~\ref{fig:successful predtions1} and \ref{fig:successful predtions2}. The first row represents 5 frames out of 100 frames of one reference sample. The second row illustrates the prediction corresponding to the frames in the first row, and the last row represents the error in the corresponding predictions. The columns represent the time steps 1, 25, 50, 75 and 100 seconds. We visualized frames at intervals of 25 seconds to evaluate different ranges of dynamic stress prediction.\\

For visualization purposes, the references and predictions in Figs.~\ref{fig:successful predtions1} and \ref{fig:successful predtions2} are scaled to the same range using the maximum and the minimum of each sample. The errors are scaled independently. As it can be seen in Fig.~\ref{fig:successful predtions1}, the predicted frames are quite similar to their corresponding references. Although the geometry contains sharp corners and edges, which are areas that are hard for CNN to reconstruct, our model is able to predict it. The errors, except for a small part of the first frame, are in an acceptable range which shows the prediction accuracy of our model. Fig.~\ref{fig:successful predtions2} shows another successful reconstruction. Comparing references with their corresponding predicted frames demonstrates that our \ourmethod{} model can capture both load variations and maximum stress values at the same time. Furthermore, these results demonstrate that our model is able to predict a dynamic stress distribution with a high variation of distributed stress.

\begin{figure}[!h]
    \centering
    \includegraphics[width=\textwidth]{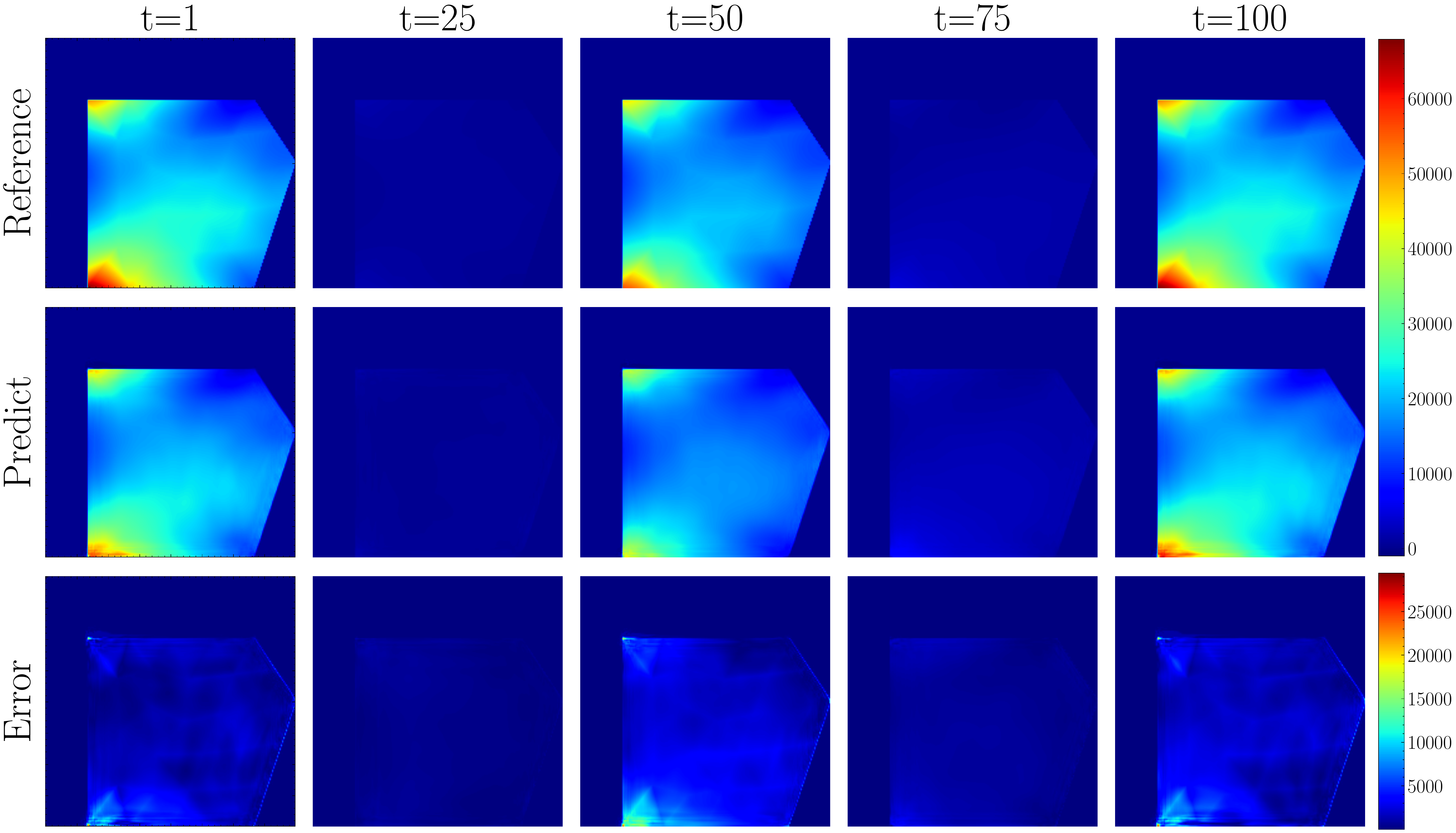}
    \caption{Failed predicted dynamic stress distribution and their corresponding errors in different time sequences. (Unit = MPa)\label{fig:failed predtions}}
\end{figure}

Fig.\ref{fig:failed predtions} shows a random failure sample. In spite of the model's success in predicting most parts of the frames, it is not able to reconstruct high-stress concentrations at angles of 90 degrees. Since CNNs typically struggle in handling sharp edges, smoothening the sharp corners using Gaussian filters during data preprocessing may help the network to train better. Furthermore, as the loads in frames $t=25$ and $t=75$ are lower than in other frames, the prediction in those frames is acceptable.\\

\begin{figure}[!h]
    \centering
   \subcaptionbox{\label{fig:successful elementwise predtions}}{\includegraphics[width=0.95\textwidth]{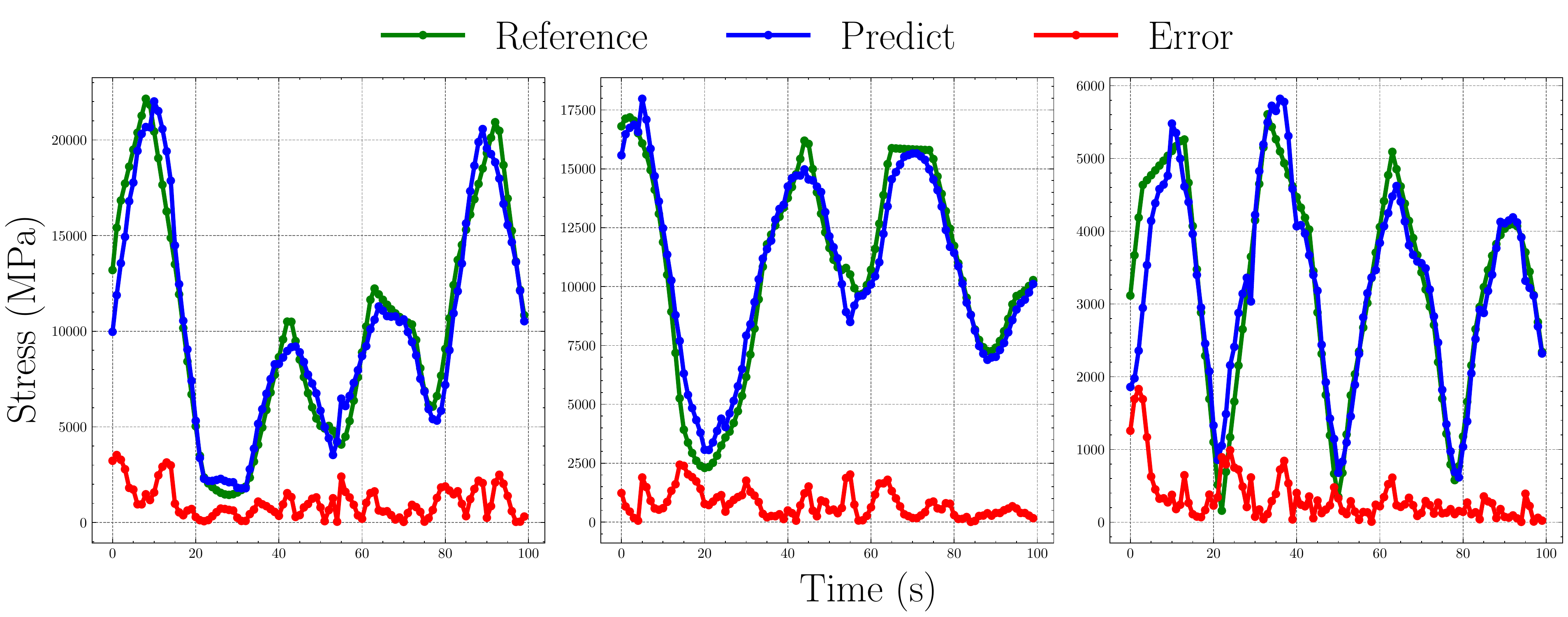}} \\
   \subcaptionbox{\label{fig:Unsuccessful elementwise predtions}}{\includegraphics[width=0.95\textwidth]{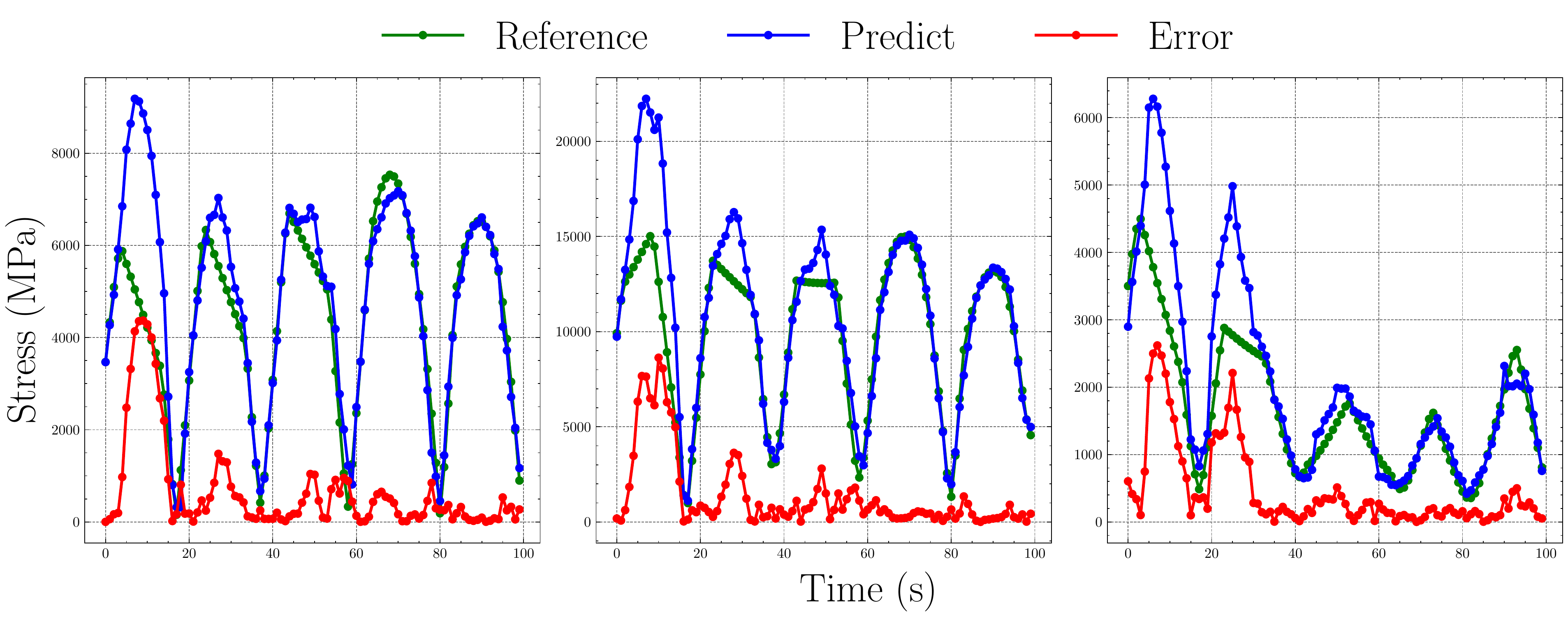}}
   \caption{{Comparison of stress values across 100 frames for predictions, references, and errors in a randomly selected element.}  (a) Successful predictions (b) Unsuccessful predictions  (Units = MPa-T).}
\end{figure}

It is also important that the predictions are temporally consistent. In order to qualitatively demonstrate the temporal consistency of the proposed method, Fig.~\ref{fig:successful elementwise predtions} shows a comparison of stress values across 100 frames for successful predictions in a randomly selected element. As can be seen, the references and the predicted distributions are almost identical in most time sequences, with errors close to zero, despite the stress varying widely with time. Fig.~\ref{fig:successful elementwise predtions} illustrates how prediction fits with reference more closely when there is more temporal smoothness at peak points. For instance, a good match between prediction and reference can be seen in the rightmost graph in Fig.~\ref{fig:successful elementwise predtions}, where the stress variation follows a smooth Gaussian distribution in the last peak. However, in the remaining graphs, the prediction has good correlation with the reference despite a lack of smoothness in most peak stress values. Moreover, based on the graphs in Fig.~\ref{fig:successful elementwise predtions}, we can conclude that the model is better at predicting stress in valleys compared to peaks.\\

\begin{figure}[!h]
    \centering
   \subcaptionbox{\label{fig:elementwise MRPE 1}}{\includegraphics[width=0.95\textwidth]{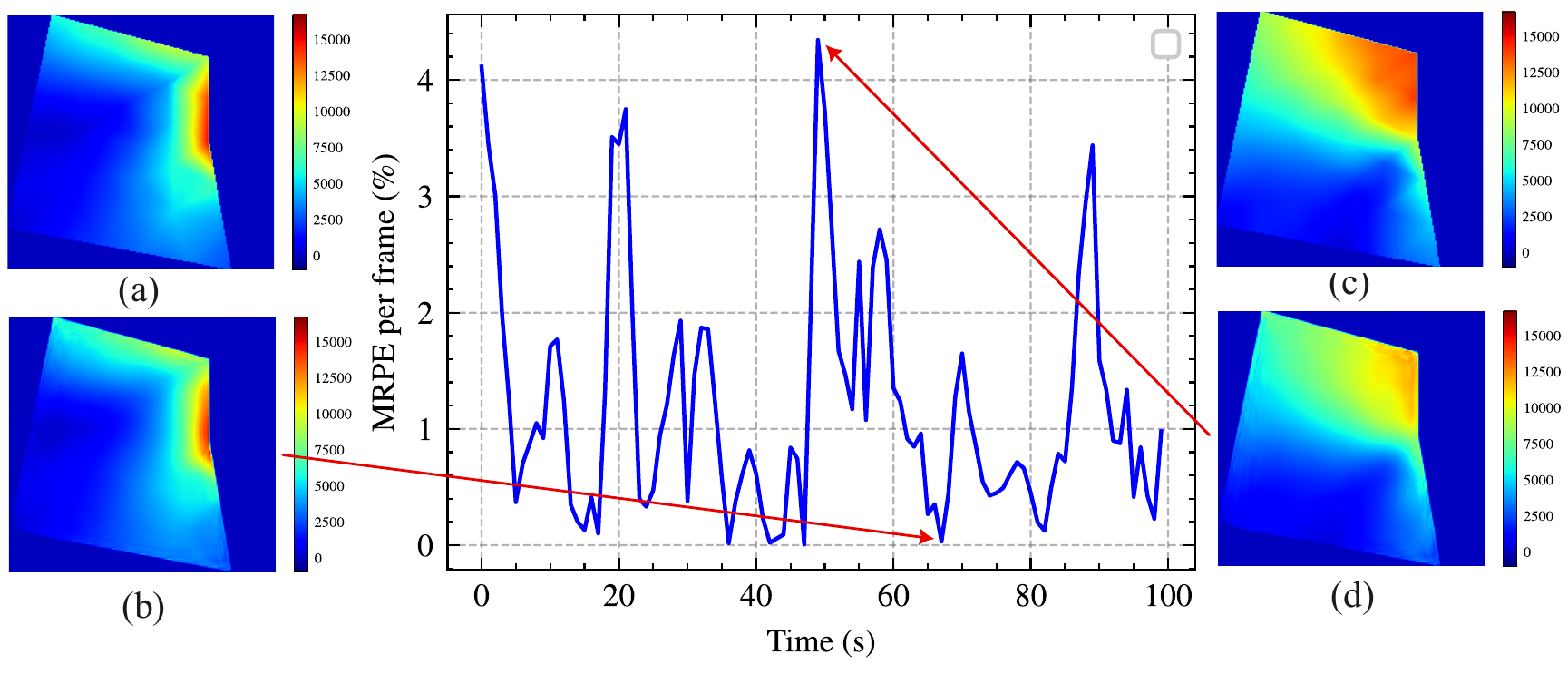}} \\
   \subcaptionbox{\label{fig:elementwise MRPE 2}}{\includegraphics[width=0.95\textwidth]{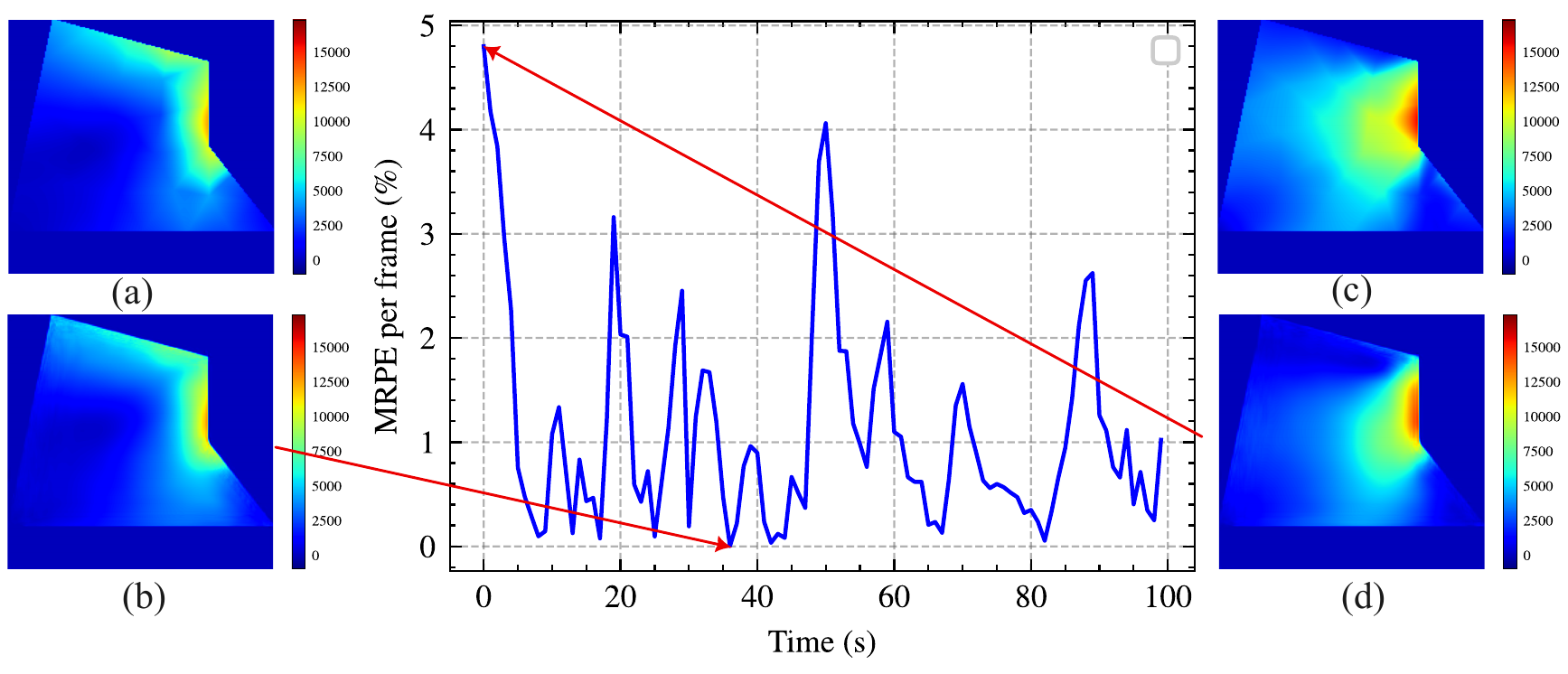}}
   \caption{{Relative errors across 100 frames in the randomly selected sample.} Graphs in the center represent the MRPE per frame.  (a) and (c) in each figure represent the reference; (b) and (d) refer to their corresponding predictions. Arrows refer to the MRPE of the presented frame. (Units = MPa-T).}
\end{figure}

We have also illustrated some of the unsuccessful predictions in Fig.~\ref{fig:Unsuccessful elementwise predtions} to identify the limitations of our proposed model. It can be seen that in all graphs with non-Gaussian stress distributions, the model finds it difficult to capture the peak stress values accurately. However, in the first two graphs from the left in Fig.~\ref{fig:Unsuccessful elementwise predtions}, the predictions perfectly fit later peaks of the reference since the stress values in the reference have Gaussian distributions at these points. Figs.~\ref{fig:elementwise MRPE 1} and \ref{fig:elementwise MRPE 2} depict the MRPE of randomly selected samples across 100 frames and frames corresponding to the minimum and maximum MRPE. As can be seen for both samples, the minimum errors are around zero, with only a few frames exceeding the error by more than 2\%.

\subsection{Ablation Study}

The efficiency of architecture can be attributed to several design choices we have made. Our architecture models the temporal dependency between time frames and the relationship between different elements in an input. Even though self-attention has shown state-of-the-art performance in sequence modeling, they are not suitable for tasks without large amounts of data. Hence, we use LSTMs for sequence modeling. To demonstrate our claim, we compare our architecture against other baseline architectures. We compare against three architectures as shown in Table~\ref{table:arch-comparison}. The model with multi head self-attention is very similar to our architecture, except the LSTM modules in our model are replaced with self-attention modules. The details of the other models are represented in Table~\ref{table:arch-comparison}. We will refer to our architecture as \ourmethod{}.  The results are shown in Table~\ref{table:arch-comparison}, and the best results are highlighted in bold.

\begin{table}[!h]
\begin{center}
\caption{Architecture comparison}
\label{table:arch-comparison}
\begin{tabular}{p{3cm} >{\centering}p{4cm} >{\centering}p{1.5cm} >{\centering}p{1.5cm} >{\centering\arraybackslash}p{1.5cm}}
\toprule
\multicolumn{5}{c}{Architecture for modeling temporal information} \\
\hline
 & Multi-headed self-attention & LSTM & LSTM  & LSTM \\
\noalign{\smallskip}
\midrule
 FaPN  & \checkmark & \checkmark &\checkmark &  $\times$ \\
 Skip connection  &\checkmark & \checkmark & $\times$ & $\times$ \\
 \midrule
 MRPE(\%) & 4.5 & \textbf{2.3} & 6.6 & 9.7  \\
\noalign{\smallskip}
\bottomrule
\end{tabular}
\end{center}
\end{table}

\section{Conclusion}
We propose \ourmethod{} model equipped with Convolutional Neural Network~(CNN) and Long Short Term Memory~(LSTM) to predict the entire sequence of dynamic stress distribution. The model was designed and trained to use the geometry, boundary conditions and the sequence of loads as input and predicts the sequence of high-resolution dynamic stress contours. The convolutional components are used to extract spatial features and the LSTM captures the temporal dependence between the frames. Feature alignment modules are used to improve the training and performance of our model. The model is trained using synthetic data generated using the PDE toolbox in MATLAB. \ourmethod{} can predict dynamic stress distribution with a mean relative percentage error of 2.3\%, which is considered an acceptable error rate in engineering communities.

\section*{Declarations}

\begin{itemize}
\item This research was funded in part by the National Science Foundation grant CNS 1645783.
\item There is no conflict of interest among the authors of this paper
\item The datasets generated during and/or analyzed during the current study are available from the corresponding author upon reasonable request.
\end{itemize}

\bibliography{main}


\end{document}